\documentclass[a4paper]{jpconf}
\usepackage{graphicx}

\newcommand{\be}{\begin{equation}}
\newcommand{\ee}{\end{equation}}
\newcommand{\bea}{\begin{eqnarray}}
\newcommand{\nn}{\nonumber}
\newcommand{\eea}{\end{eqnarray}}

\begin{document}
\title{Multipole Moments of numerical spacetimes}

\author{George Pappas$^{1,2}\dag$ and Theocharis A Apostolatos$^1\ddag$}

\address{$^1$~Section of Astrophysics, Astronomy, and
Mechanics, Department of Physics, University of Athens,
Panepistimiopolis Zografos GR15783, Athens, Greece}
\address{$^2$~Theoretical Astrophysics, IAAT, Eberhard Karls University of T\"ubingen, T\"ubingen 72076, Germany}

\ead{$\dag$~georgios.pappas@guest.uni-tuebingen.de}
\ead{$\ddag$~thapostol@phys.uoa.gr}

\begin{abstract}
In this article we present some recent results on identifying correctly the relativistic multipole moments of
numerically constructed spacetimes, and the consequences that this correction has on searching for appropriate analytic
spacetimes that can approximate well the previously mentioned numerical spacetimes. We also present expressions that give the
quadrupole and the spin octupole as functions of the spin parameter of a neutron star for various equations of state
and in a range of masses for every equation of state used. These results are relevant for describing the exterior spacetime
of rotating neutron stars that are made up of matter obeying realistic equations of state.
\end{abstract}

\section{Introduction}

The multipolar expansion of the gravitational field in Newtonian gravity is a straightforward exercise.
In spherical coordinates one could express the field as an expansion in powers of the radial coordinate and
the appropriate angular eigenfunctions of the Laplace operator. One could be tempted to straightforwardly
apply the same procedure to General Relativity (GR), but in this case things are more complicated due to the
non-linear nature of the theory (the principle of superposition does not apply in GR). The task of defining the
multipole moments in GR was undertaken in the beginning of the 70s by Geroch and Hansen who defined the multipole
moments of an asymptotically flat spacetime in the static and stationary case as tensors at infinity \cite{Gero,Hans}.
Alternative definitions of the relativistic multipole moments were also given in the early 80s by Simon \cite{Simon}
and by Thorne \cite{Thorne} which were closer to the spirit of the Newtonian asymptotic expansion, where Simon's moments
end up to be the same moments as Geroch's and Hansen's, while Thorne's moments are coordinate dependent and an
appropriate choice of the coordinates should be made in order to read the usual Geroch-Hansen moments
(for a review see \cite{Quevedo}). Finally in 1989, Fodor et al.~\cite{Fodoretal} developed an algorithm for computing
the multipole moments of a spacetime that is additionally axially symmetric, taking advantage of the insightful
Ernst-potential formalism, expressing thus the scalar (in this case) multipole moments in terms of the Ernst potential.

In the case of stationary and axially symmetric spacetimes, the relativistic multipole moments can fully characterize the
spacetime as one can see in \cite{Fodoretal}. Thus the multipole moments are of interest as quantities that
capture all the properties of a given spacetime. On the other hand, they are also of interest on experimental grounds,
since they can be measured in gravitational wave signals from gravitational inspirals. In particular, Ryan \cite{Ryan}
related various observed properties of these inspirals to the multipole moments of the background spacetime. Specifically
he gave relations that connected the multipole moments of a spacetime to precession frequencies of orbits on that
spacetime that deviate slightly from being circular equatorial ones as well as to the number of cycles
of the gravitational wave signal that is emitted by a low mass object inspiraling adiabatically into such a spacetime.
These same expressions can be also used in the context of accretion discs and the quasi-periodic oscillations (QPOs)
observed in the X-ray spectrum of low mass X-ray binaries (LMXRBs) \cite{derKlis}, where one, assuming a relativistic
precession model for the QPOs \cite{stella1}, can relate the QPO frequencies to the multipole moments of the central
object \cite{pappasQPOs}. The connection of the multipole moments to the structure and the physical properties of a
compact object, like a neutron star, was first attempted by Laarakkers and Poisson, who related the multipole moments
of a neutron star to the equation of state (EOS) for the matter in its interior \cite{LaarPois}.

Furthermore, the multipole moments can be used to construct analytic spacetimes that will have prescribed properties.
Thus, in the case that we would like to have an analytic description of the spacetime around a neutron star with specific
physical properties, we could use the appropriate multipole moments to construct such a spacetime following the
procedure presented in \cite{Sotiriou,Pachon,Pappas2,PachonNew,pappas2sol} (the multipole moments can also be important
in different approaches to the problem of constructing geometries for neutron stars, such as \cite{HT,Ruffini}).

Working on the previously mentioned applications of multipole moments to analytic and numerical spacetimes, we came to
realize that there is a problem with the way the multipole moments of numerically constructed spacetimes are
identified from their asymptotic behavior. These issues were addressed and clarified in \cite{PappasMoments}
where the correct expressions for the quadrupole and the spin octupole were derived using Ryan's \cite{Ryan}
algorithm for relating the gravitational-wave spectrum $\Delta \tilde{E}$ (the energy emitted per unit logarithmic
frequency interval) of a test particle that is orbiting on a circular equatorial orbit in
an asymptotically flat, stationary and axially symmetric spacetime, to the multipole moments of that spacetime.
In \cite{PappasMoments} various consequences of correcting the calculation of the numerical multipole moments
were also discussed, such as the effect this correction has on attempting to approximate the numerical spacetime
exterior to neutron stars. In particular it was shown that when one uses an analytic spacetime that depends on a number
of parameters that can be connected to the multipole moments of a numerical spacetime, such as the analytic
spacetime of Manko et al. \cite{manko1,manko2}, in order to approximate that numerical spacetime, the use of the
corrected values for the multipole moments improved the performance of the analytic spacetime and resulted to a better fit
to the numerical spacetime.

It was also shown, extending a previous result by Laarakkers and Poisson \cite{LaarPois}, that the reduced quadrupole,
$q\equiv M_2/M^3$, and the reduced spin octupole, $s_3\equiv S_3/M^4$, follow respectively
quadratic and cubic dependance on the spin parameter, $j=J/M^2$, just as in the case of the Kerr spacetime
where these moments are $-j^2$ and $-j^3$ respectively, with the difference being that the proportionality constant is
larger than 1 for neutron stars.

In this work we briefly present these results, i.e., we give a short derivation of the correct expressions for
the multipole moments and then present the resulting improvement in the fit of the numerical metric by the 3-parameter
analytic metric of Manko et al. In addition we present some results from approximating the numerical metric
with the 4-parameter two-soliton metric, that is extensively discussed in \cite{pappas2sol}, that further strengthen our
arguments in favor of using the corrected multipole moments as criteria for matching an analytic spacetime to
a numerical one. Finally we give the proportionality constants between the quadrupole and the square of
the spin parameter, $j^2$, and the spin octupole and the cube of the spin parameter, $j^3$, for different equations
of state and for different masses. These coefficients are directly comparable to the coefficients of Laarakkers and Poisson
for the quadrupole. All the numerical spacetimes are calculated using the RNS numerical code \cite{SterFrie}. All the
expressions are in units of $G=c=1$.

\section{\label{sec_2}Asymptotic expansion of a metric and the multipole moments}


The line element used for the spacetime for the interior and the exterior of a neutron star, which has symmetry
with respect to time translations and rotations, i.e., it is stationary and axisymmetric,  is usually written
in quasi-isotropic coordinates in the form,

\be\label{isotropic}
ds^2 = -e^{2\nu} dt^2 + r^2 \sin^2\theta B^2 e^{-2\nu}(d\phi-\omega dt)^2 + e^{2(\zeta-\nu)} (dr^2+r^2d\theta^2),
\ee
where $\nu$, $B$, $\omega$, and $\zeta$ are the four metric functions, all functions of the quasi-isotropic
coordinates $(r,\theta)$, as it was introduced by Butterworth and Ipser \cite{ButtIpse}. From Einstein's field
equations, one can show that the asymptotic expansion of the metric functions takes the following form,

\bea
\nu &\sim&
\left\{-\frac{M}{r}+\frac{\tilde{B}_0M}{3r^3}+\frac{J^2}{r^4}+\left[-\frac{\tilde{B}_0^2}{5}
               +\frac{\tilde{B}_2^2}{15}-\frac{12J^2}{5}\right]\frac{M}{r^5}
               +\ldots\right\}+ \left\{\frac{\tilde{\nu}_2}{r^3}+\ldots\right\}P_2
    +\ldots , \label{first}\\
\omega &\sim&
\left
[\frac{2J}{r^3}-\frac{6JM}{r^4}+\left(8-\frac{3\tilde{B}_0}{M^2}\right)\frac{6JM^2}{5r^5}
                +\ldots\right
                ]\frac{d P_1}{d\mu}+ \left[\frac{\tilde{\omega}_2}{r^5}+\ldots\right]\frac{d P_3}{d\mu}+\ldots ,
                \label{second}\eea
\be B \sim
 \sqrt{\frac{\pi}{2}}\left[\left(1+\frac{\tilde{B}_0}{r^2}\right)T_0^{1/2}
                 \left(\frac{\pi}{2}\right)^{1/2}
                 +\frac{\tilde{B}_2}{r^4}T_2^{1/2}+\ldots\right]. \label{third}
\ee
In the formulae above $P_l$ 
are the Legendre polynomials expressed as functions of $\mu=\cos\theta$, $T^{1/2}_l$ are the so called Gegenbauer
polynomials (similar to the Legendre polynomials, also functions of $\mu$), and $M,J$ are the first two
multipole moments (the mass and the spin) of the spacetime. The rest of the coefficients are related to the higher
multipole moments.

These metric functions can be rewritten using the following redefinitions,
\be
B e^{-\nu}=e^{\beta}~,~ \zeta=\nu+\alpha.
\ee
Then the combinations of $\nu$ and $\beta$,
\be
\gamma=\nu+\beta~,~\rho=\nu-\beta,
\ee
along with $\omega$ could again be expressed as power series in $1/r$ (this is done in \cite{KEH,CST,SterFrie}) in the same
manner as in Eqs.~(\ref{first},\ref{second},\ref{third}):
\bea
\rho &=& \sum_{n=0}^{\infty}(-2\frac{M_{2n}^*}{r^{2n+1}}+\textrm{higher order}) P_{2n}(\mu),\\
\omega&=&\sum_{n=1}^{\infty}(-\frac{2}{2n-1}\frac{S_{2n-1}^*}{r^{2n+1}}+\textrm{higher order})\frac{P_{2n-1}^1(\mu)}{\sin\theta},\\
\gamma&=&\sum_{n=1}^{\infty}(\frac{D_{2n-1}}{r^{2n}}+\textrm{higher order})\frac{\sin(2n-1)\theta}{\sin\theta}.
\eea
By a simple comparison between the above expansion and the corresponding ones in
Eqs.~(\ref{first},\ref{second},\ref{third}),
one can see for example that $M_{2}^*=-\tilde{\nu}_2$ and $S_3^*=\frac{3}{2} \tilde{\omega}_2$.
The coefficients $M_{2n}^*$ and $S_{2n-1}^*$ were mistakenly identified as the mass and current-mass moments, respectively,
of the corresponding spacetime and it is exactly these quantities that the numerical code of Stergioulas and Friedman
\cite{SterFrie}, RNS, provides.

In order to identify the correct multipole moments, i.e., the Geroch-Hansen multipole moments, of a spacetime given by
the line element (\ref{isotropic}), we have used Ryan's expression \cite{Ryan} that relates the multipole moments
with the energy change per logarithmic interval of the rotational frequency
$\Delta \tilde{E}=-d\tilde{E}/d \log{\Omega}$ for circular equatorial orbits in a stationary and axisymmetric spacetime.
Thus, for the line element of Eq.~(\ref{isotropic}), we first expressed the orbital frequency of circular equatorial orbits,

\be
\Omega=\frac{d\phi}{dt}=\frac{-g_{t\phi,r}+\sqrt{(g_{t\phi,r})^2-g_{tt,r}g_{\phi\phi,r}}}{g_{\phi\phi,r}},\ee
as a power series in $x=(M/r)^{1/2}$ and then inverted it to obtain,
\bea
x&=&\upsilon+\frac{\upsilon^3}{2}+\frac{j\upsilon^4}{3}+\frac{1}{24}
(13+4b-6q) \upsilon^5+\frac{j
\upsilon^6}{2}+\frac{(97+28b+56j^2-144q)\upsilon^7}{144}\nn\\
   &&+\frac{(373j+292bj-330jq-270w_2)\upsilon^8}{360}+O(\upsilon^9), \eea
where $\upsilon=(M \Omega)^{1/3},\; j=\frac{J}{M^2},\; q=\frac{\nu_2}{M^3},\;w_2=\frac{\omega_2}{M^4},\;b=\frac{B_0}{M^2}$.
Then we expanded

\be \tilde{E}=
\frac{-g_{tt}-g_{t\phi}\Omega}{\sqrt{-g_{tt}-2g_{t\phi}\Omega-g_{\phi\phi}\Omega^2}},\ee
which is the energy per unit mass for a specific circular orbit, as a power series in $x=(M/r)^{1/2}$ and substituted
the previous expression for $x$. From the resulted expression we calculated $\Delta \tilde{E}$, from the equation,
\be \Delta \tilde{E}=-\frac{d\tilde{E}}{d \log{\Omega}}=-\frac{\upsilon}{3}\frac{d\tilde{E}}{d\upsilon},\ee
and arrived to the expression,

\bea
\Delta\tilde{E} \!\!\!&=&\!\!\! \frac{\upsilon^2}{3}- \frac{\upsilon^4}{2}+ \frac{20 j \upsilon^5}{9}-
\frac{(89 +32b +24q)}{24} \upsilon^6+ \frac{28 j\upsilon^7}{3}  \nn\\
\!\!\!&-&\!\!\! \frac{5\left(1439+ 896b- 256j^2+ 672 q\right) \upsilon^8}{432}
+ \frac{\left((421+64b-60q)j-90w_2\right) \upsilon^9}{10}
+O(\upsilon^{10}).
\label{DE}
\eea
If one compares the previous expression to the one produced by Ryan,

\bea \Delta\tilde{E}\!\!\!&=&\!\!\!\frac{1}{3} \upsilon^2-\frac{1}{2}\upsilon^4 +\frac{20}{9}\frac{S_1}{M^2}\upsilon^5
       +\left(-\frac{27}{8}+\frac{M_2}{M^3}\right)\upsilon^6+\frac{28}{3}\frac{S_1}{M^2}\upsilon^7\nn\\
        \!\!\!&+&\!\!\!\left(-\frac{225}{16}+\frac{80}{27}\frac{S_1^2}{M^4}+\frac{70}{9}\frac{M_2}{M^3}\right)\upsilon^8
        +\left(\frac{81}{2}\frac{S_1}{M^2}+6\frac{S_1M_2}{M^5}-6\frac{S_3}{M^4}\right)\upsilon^9+\ldots
                 \eea
where $S_1=J$, one can see that from the coefficients of $v^6$ and $v^9$ terms of
the two series, the following values for the quadrupole and the spin octupole can be obtained:
\bea \label{correct_M2}
M_2^{GH} &=& -\tilde{\nu}_2 - \frac{4}{3} \left(\frac{1}{4}+b\right) M^3= M_2^*-\frac{4}{3}\left(\frac{1}{4}+b\right)M^3,\;
%
\\
S_3^{GH}&=& \frac{3}{2} \tilde{\omega}_2 - \frac{12}{5} \left(\frac{1}{4}+b\right) j M^4=
S_3^*-\frac{12}{5}\left(\frac{1}{4}+b\right)jM^4.
\label{correct_S3}
\eea
Henceforth we will omit the superscript $^{GH}$, that indicates the Geroch-Hansen moments, in $M_2,S_3$ when we refer
to the correct multipole moments.

Laarakkers and Poisson performed in \cite{LaarPois} a similar calculation in order to identify the quadrupole of the
metric (\ref{isotropic}), but their result was missing the last term in Eq.~(\ref{correct_M2}), i.e., the term with
$(1/4+b)$. That was because they erroneously assumed that the correct asymptotic behavior for the metric should be
that of Schwarzschild, which has $B=1-M^2/4 r^2,$ which corresponds to $b=-1/4$. Thus they had a priori fixed the
correcting last term to be zero. Under a more careful consideration though, there is no reason to fix the asymptotic
behavior of a stationary and axially symmetric spacetime in this way. A counter example for this is the case of the Kerr
spacetime, where one can see that in quasi-isotropic coordinates the metric function $B$ is given by,

\be B_{{\tiny\textrm{Kerr}}}=1-\frac{(M^2-a^2)}{4 r^2},\ee
which corresponds to having $b^{{\tiny\textrm{Kerr}}}=-(1/4)(1-j^2)$.

\section{\label{sec_3}Improvement in approximating a numerical spacetime with analytic spacetimes due to the
 correction in the moments}


We will now present the effect that the correction in the moments has when one attempts to use an analytic spacetime
to approximate the exterior space time of a neutron star, that is calculated numerically, by matching
the multipole moments of the analytic spacetime to the multipole moments of the numerical spacetime. Berti and Stergioulas
\cite{BertSter} tried to match a three-parameter analytic solution, the solution of Manko et al. \cite{manko1}, to a wide
diversity of uniformly rotating neutron-star models. Each analytic solution was constructed so that its first three
multipole moments were equal to the corresponding moments (mass, spin and quadrupole) of the particular numerical neutron
star, where these moments were read from the corresponding numerical metric. For their calculations, they used as moments
the quantities $M_{2n}^*$ and $S_{2n-1}^*$. Their conclusion was that this type of analytic solution was quite good
to describe the external metric of all kinds of fast rotating neutron stars. Since the specific metric cannot assume
low values of quadrupole moment, the metric is not adequate to describe rotating neutron stars with rotation lower than
some value. We have used the Manko et al. solution to demonstrate the effect of correcting the multipole moments,
compared to the results by Berti and Stergioulas.

Since the Manko et al. solution is in the previously mentioned sense
handicapped, and since it has only three free parameters which provide the freedom to explore the behavior of the
metric by changing only the quadrupole, we have also used another analytic solution, the so called two-soliton solution
\cite{twosoliton}, that can also be used to approximate the exterior spacetime of neutron stars \cite{pappas2sol} and
has four parameters, which allow into play the spin octupole as well. Thus, with the freedom now to vary both
the quadrupole and the spin octupole, we have performed comparisons between the two-soliton and specific numerical
metrics in order to see which values of multipole moments give the best fit.

In both cases, for the Manko et al. and the two-soliton, we have found that when the parameters of the analytic
metric are calculated using the correct values for the multipole moments, the fitting of the analytic metric to the
corresponding numerical one is better. This result also supports our claim that the way that an analytic metric should be
matched to the numerical metric is by identifying the corresponding multipole moments of the two spacetimes.

\begin{figure}
\includegraphics[width=0.5\textwidth]{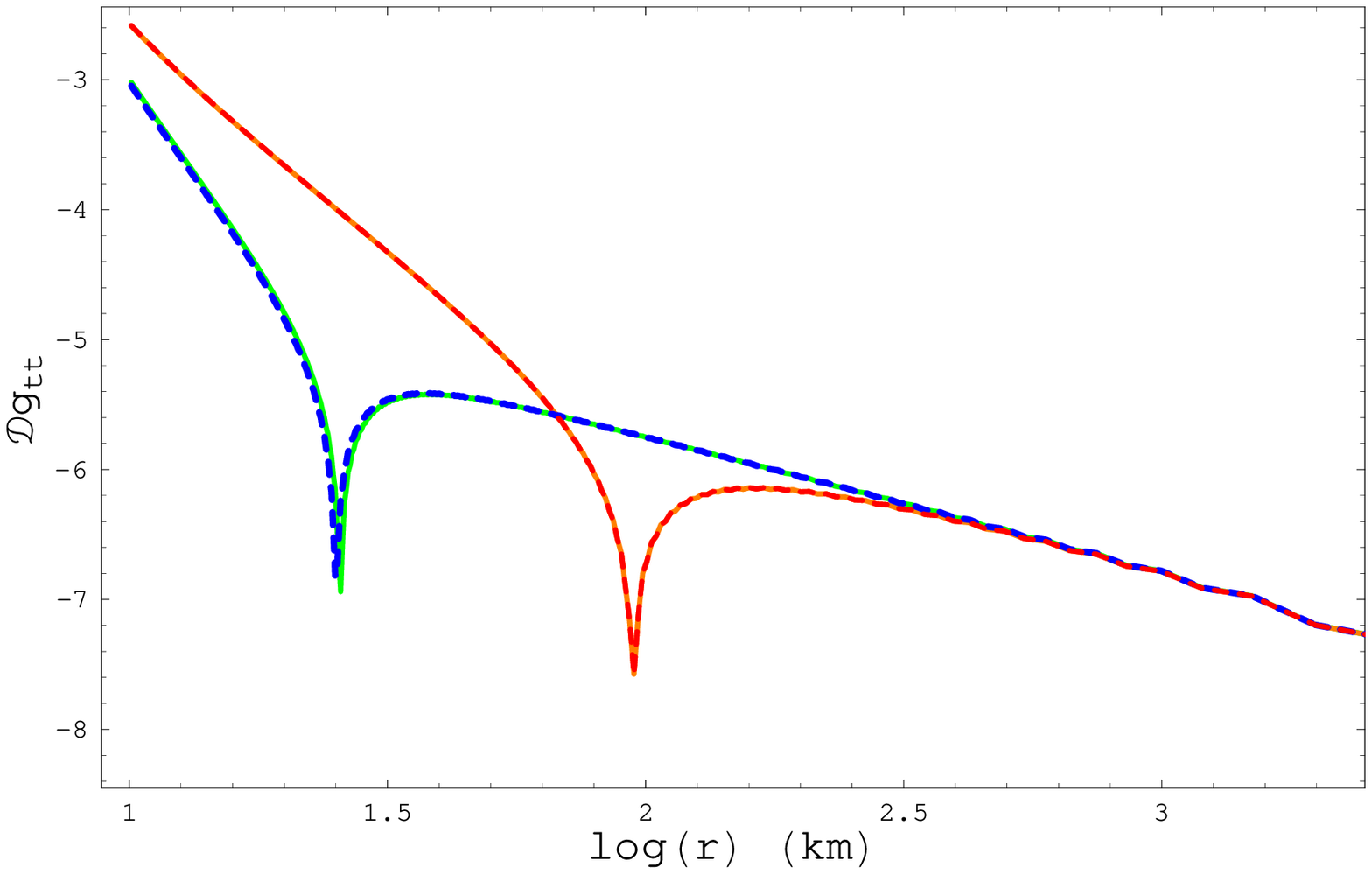}
\includegraphics[width=0.5\textwidth]{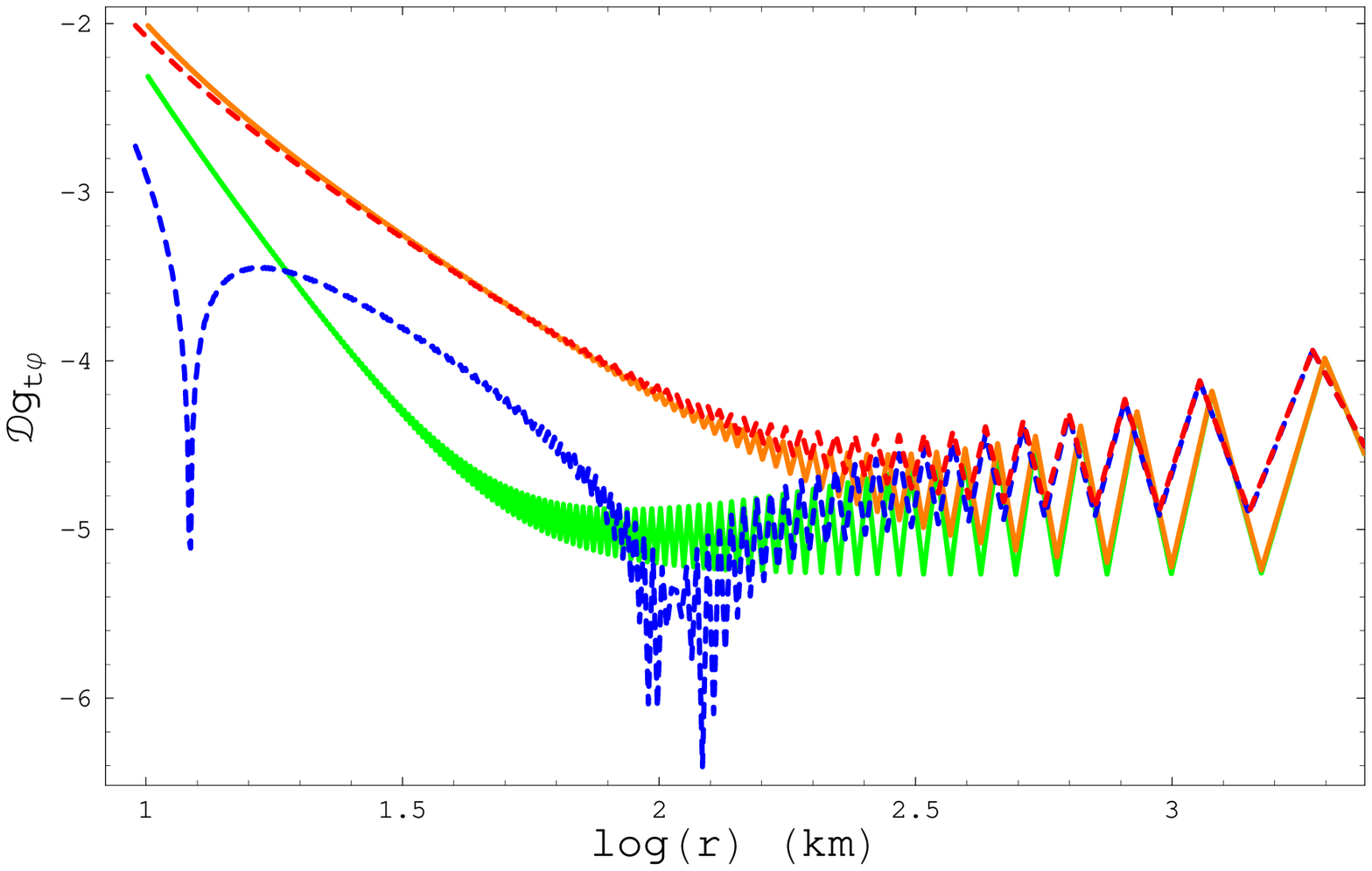}
\protect
\caption{A typical log-log plot of the relative difference between the
numerical and the analytic metric ($(g_{ij}^n-g_{ij}^a)/g_{ij}^n$) for a specific numerical model
(model $\# 19$ of EOS FPS of \cite{PappasMoments}). The left plot is for $g_{tt}$ and the right one
for $g_{t \phi}$. The red dashed curves correspond to the Manko et al. metric produced from $M_2^*$ and
the blue dashed curves correspond to the Manko et al. produced from $M_2$. The orange solid curves correspond to the
two-soliton produced using the moments $M_2^*, S_3^*$ and the green solid curves correspond to the two-soliton
produced using the corrected moments $M_2, S_3$.}
\label{fig1}
\end{figure}

In Figure \ref{fig1} we present a typical comparison between the analytic metrics that are calculated using the correct
moments, i.e. $M_2, S_3$, and the analytic metrics calculated using the previously assumed moments. Specifically the
figures show the relative difference between the numerical and the analytic metric ($(g_{ij}^n-g_{ij}^a)/g_{ij}^n$) as a
function of coordinate radius for the Manko et al. analytic metric and the two-soliton analytic metric, calculated for the
values $M_2^*, S_3^*$ and for the values $M_2, S_3$ (the value of the spin octupole is relevant only for the two-soliton
spacetime which has 4 parameters). The red dashed curves correspond to the Manko et al. metric produced from $M_2^*$ and
the blue dashed curves correspond to the Manko et al. produced from $M_2$. The orange solid curves correspond to the
two-soliton produced using the moments $M_2^*, S_3^*$ and the green solid curves correspond to the two-soliton
produced using the corrected moments $M_2, S_3$. The left plot in Figure \ref{fig1} shows the relative difference
for the $g_{tt}$ metric component while the right plot shows the relative difference for the $g_{t\phi}$ metric components.

\begin{figure}
\centering
\includegraphics[width=.32\textwidth]{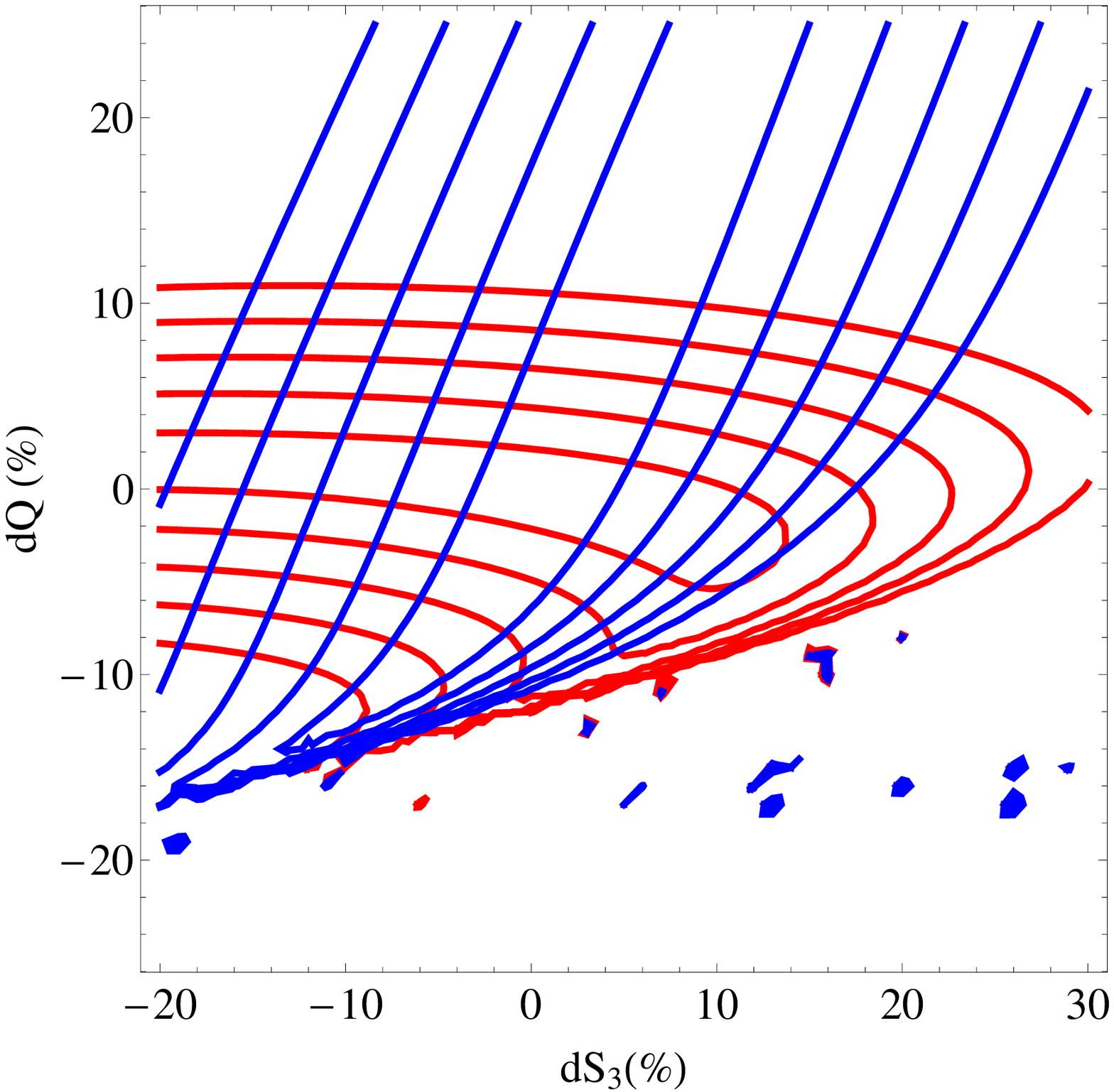}
\includegraphics[width=.32\textwidth]{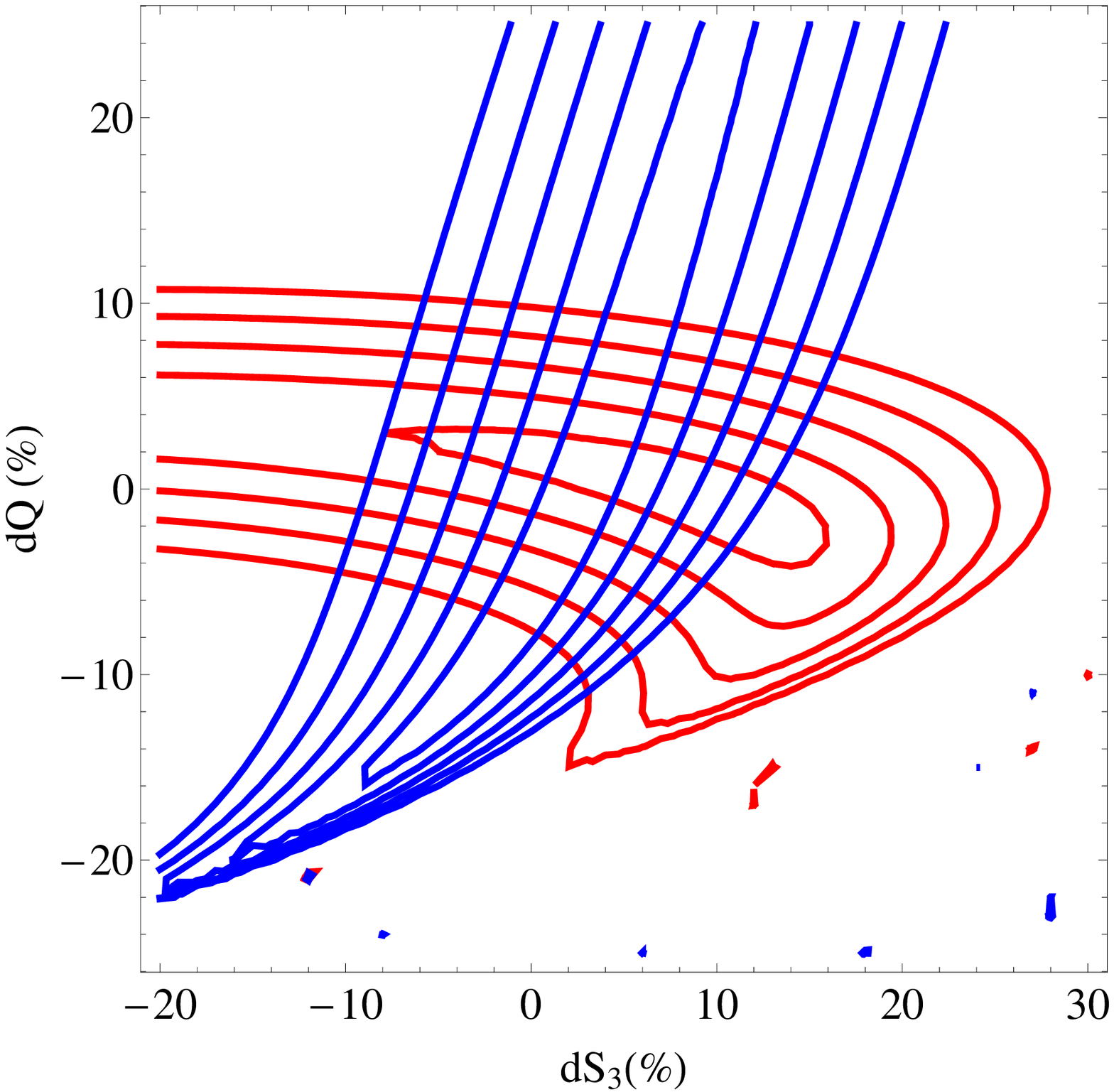}
\includegraphics[width=.32\textwidth]{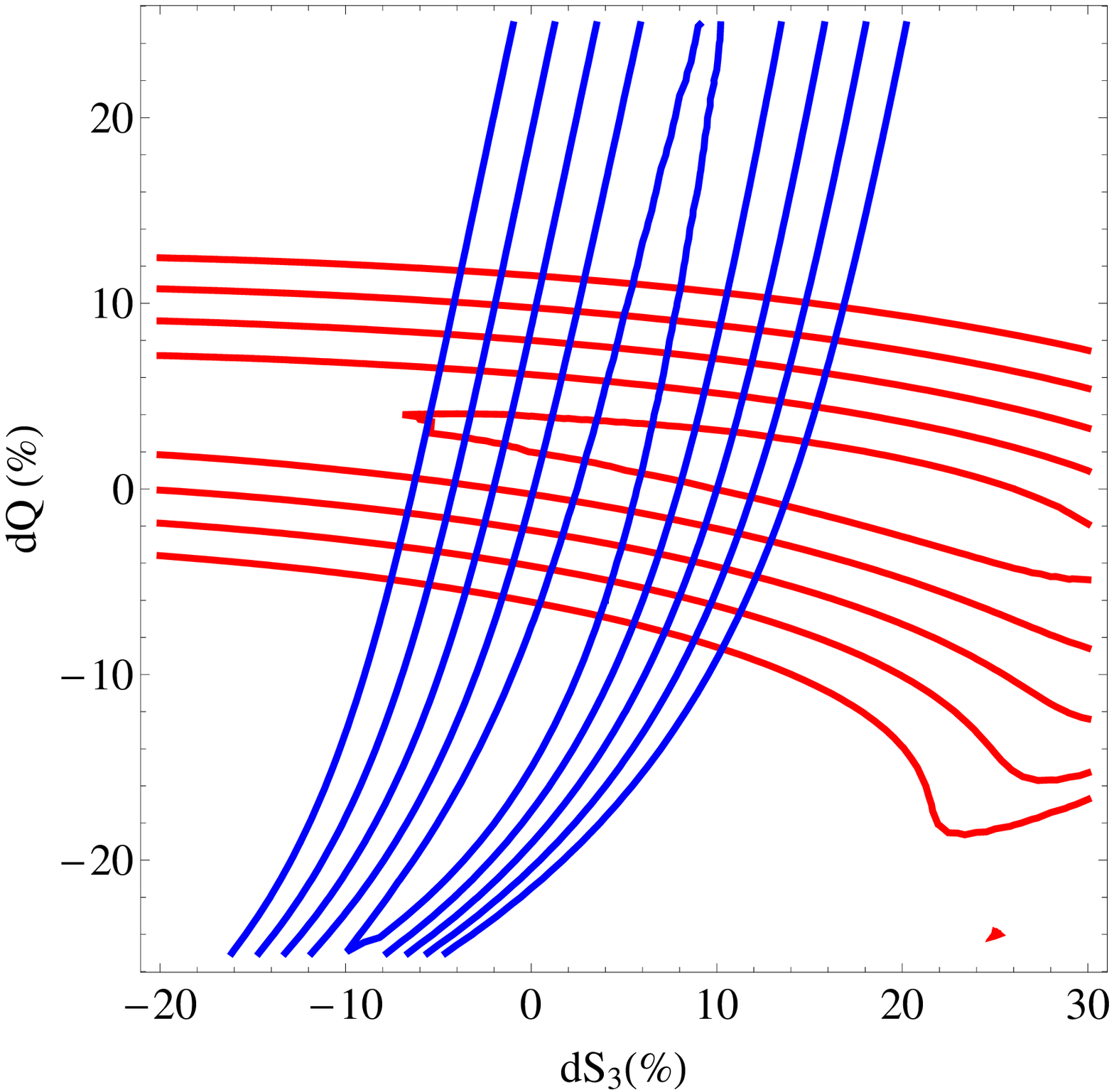}
\caption{Contour plots that demonstrate that the appropriate moments to be used for matching an
analytic spacetime to a numerical spacetime are the corrected moments $M_2, S_3$. The
plots show the contour plots of the overall mismatch $\sigma_{ij}$ between the analytic and the numerical
metric for the $tt$ (red curves) and $t\phi$ components  (blue curves) as a function of the
fractional deviation of the quadrupole, $d M_2$, and the current octupole, $d S_3$, of the analytic metric
from those calculated from the numerical metric as they are given by the corrected expressions (assuming the same mass
and angular momentum). The value of the overall mismatch increases from the inner to the outer contours.
Because the contours of the $\sigma_{tt}$ (almost horizontal) are orthogonal to the ones
of $\sigma_{t\phi}$ (almost vertical), the combination indicates an optimum choice for the multipole
moments of the analytic space-time. That choice is the moments that
have zero deviation from the moments of the numerical space-time.
These contour plots correspond, from left to right, to the models $\# 22$ and  $\# 28$ for the AU EOS and $\# 28$
for the L EOS of the models presented in \cite{PappasMoments}.}
\label{matching}
\end{figure}

In order to further test the effect that varying the moments $M_2$ and $S_3$ has on the performance of the analytic metric
to approximate the numerical metric, we have used the two-soliton metric to investigate the parameter space of the
variation of the moments. As a measure of the ability of the analytic metric to approximate the numerical metric we
have used the quantity, that we call, ``overall mismatch'' between the analytic and the numerical metric functions,
that is defined (see \cite{PappasMoments}) as
\be
\sigma_{ij}=\left[ \int_{R_{\rm S}}^{\infty}(g_{ij}^n-g_{ij}^a)^2dr \right]^{1/2},
\label{sigmamis}
\ee
where $R_{\rm S}$ is the radius $r$ at the surface of the star. For each numerical model of the various models of
uniformly rotating neutron stars that we have constructed, we formed a set of two-soliton spacetimes
that have the same mass $M$ and angular momentum $J$ with the numerical model, but the quadrupoles and the current
octupoles take the values $M_2^{\rm (a) }=M_2(1-d M_2 )$ and
 $S_3^{\rm (a) } =S_3(1-d S_3 )$ respectively, where the $d M_2$ and $d S_3$ take various values.
The quantities $d M_2$ and $d S_3$ denote the fractional differences of the corresponding moments used in the
calculation of each two-soliton spacetime from the moments of the numerical spacetime.
For each one of the two-soliton spacetimes belonging to the previously mentioned set, we calculated the ``overall
mismatch'' between the analytic and the numerical metric functions producing contour plots of the ``overall
mismatch'' on the space of $d M_2$ and $d S_3$ for the different numerical models
(one can find further discussion on this in \cite{pappas2sol}). Examples of these contour plots are shown in
Figure \ref{matching}.

These contour plots show that the choice of multipole moments for an analytic spacetime that give the best approximation
to the numerical spacetime, are those that correspond to the moments given by the equations
(\ref{correct_M2},{\ref{correct_S3}), i.e., those with $(dM_2,dS_3)\simeq(0,0)$.

\section{Relation between the multipole moments and the spin parameter $j$}

The connection of the higher moments of a compact object with its spin parameter $j$ was first attempted by Laarakkers
and Poisson \cite{LaarPois}. Specifically what they showed was that the reduced quadrupole moment of neutron stars
constructed using realistic equations of state was proportional to the square of the spin parameter and the
proportionality constant depended on the mass of the neutron star and the equation of state, i.e.,

\be q \simeq -a(M,\textrm{EOS}) j^2. \ee

In \cite{PappasMoments} it was shown that this relation is true, even after the correction of the quadrupole moment
because the correcting factor $(1/4+b)$ is proportional to $j^2$, which is strikingly similar to the way that this factor
behaves in the case of the Kerr spacetime. A similar result was shown to apply for the spin octupole as well, i.e., the
spin octupole has a dependence on the cube of the spin parameter of the form,

\be s_3 \simeq -\beta(M,\textrm{EOS}) j^3. \ee

The calculations performed in \cite{PappasMoments}, were performed using evolutionary sequences, so these results
were not directly comparable to the results of Laarakkers and Poisson. Here we perform these fits for the reduced
quadrupole and the reduced spin octupole as functions of $j^2$ and $j^3$ respectively and calculate the coefficients
$a$ and $\beta$ for neutron star models of constant mass and varying rotation. The results are presented in Table
\ref{label}.

\begin{table}[h]
\caption{\label{label}The table gives the coefficients $a, \beta$ of $q=-aj^2$ and $s_3=-\beta j^3$ for various equations
of state ranging from soft (on the left) to stiff (on the right) and for different masses in the range of
$0.9-2.1M_{\odot}$. More details on the equations of state used can be found in \cite{CST,BertSter}}
\begin{center}
\begin{tabular}{crlrlrlrlrlrl}
\br
& \multicolumn{2}{c}{A} & \multicolumn{2}{c}{AU} & \multicolumn{2}{c}{FPS} & \multicolumn{2}{c}{APR} & \multicolumn{2}{c}{UU} & \multicolumn{2}{c}{L}\\
\br
$M/M_{\odot}$ & $a$ & $\beta$ & $a$ & $\beta$ & $a$ & $\beta$ & $a$ & $\beta$ & $a$ & $\beta$ & $a$ & $\beta$\\
\mr
 0.9 & 7.80 & 16.2 & 8.23 & 17.2 & 9.08 & 19.6 & 10.9 & 23.8 & 9.27 & 19.6 & 14.1 & 30.3 \\
 1.0 & 6.50 & 13.5 & 6.86 & 14.4 & 7.82 & 16.6 & 9.39 & 20.1 & 7.73 & 16.4 & 12.2 & 26.4 \\
 1.1 & 5.56 & 11.3 & 6.03 & 12.3 & 6.69 & 14.0 & 7.67 & 16.5 & 6.81 & 14.1 & 10.9 & 23.2 \\
 1.2 & 4.69 &  9.31 & 5.17 & 10.5 & 5.85 & 12.2 & 6.81 & 14.3 & 5.87 & 12.1 & 9.37 & 19.8 \\
 1.3 & 4.07 &  8.03 & 4.47 & 8.80  & 5.05 & 10.3 & 5.98 & 12.5 & 5.13 & 10.3 & 8.76 & 18.2  \\
 1.4 & 3.49 &  6.70 & 3.99 & 7.80  & 4.33 & 8.66  & 5.21 & 10.6 & 4.59 & 9.22 & 7.60 & 15.8  \\
 1.5 & 2.97 &  5.54 & 3.50 & 6.62  & 3.86 & 7.68  & 4.69 & 9.55 & 3.98 & 7.78 & 6.69 & 13.8 \\
 1.6 & 2.58 &  4.86 & 3.14 & 5.88  & 3.33 & 6.48  & 4.13 & 8.14 & 3.60 & 6.98 & 6.23 & 12.7 \\
 1.7 &  -    &  -   & 2.74 & 4.92  & 2.92 & 5.51  & 3.80 & 7.47 & 3.12 & 5.83 & 5.55 & 11.2 \\
 1.8 &  -    &  -   & 2.52 & 4.47  & 2.49 & 4.61  & 3.32 & 6.29 & 2.73 & 4.94 & 5.06 & 10.2 \\
 1.9 &  -    &  -   & 2.23 & 3.79  &    -  &   -  & 3.07 & 5.80 & 2.55 & 4.56 & 4.63 & 9.15 \\
 2.0 &  -    &  -   & 2.07 & 3.45  &    -  &   -  & 2.70 & 4.92 & 2.24 & 3.84 & 4.33 & 8.53 \\
 2.1 &  -    &  -   & 1.78 & 2.82  &    -  &   -  & 2.40 & 4.18 & 2.08 & 3.53 & 3.93 & 7.69 \\
\br
\end{tabular}
\end{center}
\end{table}

These results are comparable to those presented by
Laarakkers and Poisson in Table VII of \cite{LaarPois}. So, we can compare the quadrupole coefficients for the
equations of state FPS and L that are given both here and in
\cite{LaarPois}. We should first note that there is a typo in Table VII that shows the fit parameters. The value of
$a$ for the $1.4M_{\odot}$ models for EOS FPS should be $4.2$. To compare the results in our Table \ref{label} and the
results in Table VII of \cite{LaarPois} we will calculate the relative difference between the quadrupole moments one would calculate
from our coefficients and from the coefficients in \cite{LaarPois}. The relative difference in the quadrupole moments
would be

\be \Delta M_2=\frac{M_2^{{\tiny\textrm{LP}}}-M_2}{M_2}
       =\frac{(-a^{{\tiny\textrm{LP}}}j^2M^3)-(-a j^2M^3)}{(-a j^2M^3)}=\frac{a^{{\tiny\textrm{LP}}}-a}{a}, \ee
i.e., equal to the relative difference in the coefficients. The results are shown in Table \ref{Tab2}, where we see that
there is an increase in the error of calculating the quadrupole as we go to models of higher mass, that for the FPS EOS
can be as high as 11\% for the models of $1.8M_{\odot}$.

\begin{table}[h]
\caption{\label{Tab2}The table gives the relative differences in the coefficients $a$ for the quadrupole between
the results presented in Table \ref{label} and in Table VII of \cite{LaarPois} for the equations of state
FPS and L. The relative differences are given as \%.}
\begin{center}
\begin{tabular}{c|ccccc}
\br
\br
$M/M_{\odot}$ & 1.0 & 1.2 & 1.4 & 1.6 & 1.8 \\
\mr
 EOS FPS & -0.26 & -2.56 & -3.00 & -6.91 & -11.65 \\
 EOS L   & -0.82 & -0.75 & -2.63 & -3.69 & -3.16 \\
\br
\end{tabular}
\end{center}
\end{table}

\section{Conclusions}

In this work we have briefly presented some recent results (see \cite{PappasMoments}) regarding the correct
identification of the multipole moments of a spacetime given in quasi-isotropic coordinates, with application to
the evaluation of the multipole moments of numerical spacetimes. We have also briefly discussed the effect that this
correction has on using the multipole moments as parameters for constructing parameterized analytic spacetimes
that approximate numerical spacetimes around compact objects. Specifically we showed that when the correct moments are
taken into account, an analytic spacetime gives a better approximation of the numerical spacetime and further more,
for the case of the two-soliton spacetime, the best approximation is the one that implements the corrected
quadrupole and spin octupole. Finally we have discussed the fact that the quadrupole and the spin octupole
seem to be proportional to the square of the spin parameter, $j^2$, and the cube of the spin parameter, $j^3$,
respectively, and we have presented the proportionality constants for various equations of state and various masses.

We hope that the coefficients presented in Table \ref{label} will be useful to anyone that would like to
construct parameterized analytic spacetimes for the exterior of neutron stars, that involve the first four non-zero
multipole moments or relations between them as parameters.

\ack
We would  like to thank Kostas Kokkotas for the hospitality at the University of T\"ubingen
and Nikos Stergioulas for providing us access to his RNS numerical code. This work has been supported by the
I.K.Y. (IKYDA 2010). G.P. would also like to acknowledge DAAD scholarship number A/12/71258.

\smallskip


\section*{References}
\begin{thebibliography}{9}


\bibitem{Gero} Geroch R 1970 {\it J. Math. Phys.} {\bf 11} 1955-1961;
 Geroch R 1970 {\it J. Math. Phys.} {\bf 11} 2580-2588


\bibitem{Hans} Hansen,~R~O 1974 {\it J. Math. Phys.} {\bf 15}  46-52



\bibitem{Simon} Simon W and Beig R 1983 {\it J. Math. Phys.} {\bf 24} 1163-1171


\bibitem{Thorne} Thorne K S 1980 {\it Rev. Mod. Phys.} {\bf 52} 299-399


\bibitem{Quevedo} Quevedo H 1990 {\it Fortschr. Phys.} {\bf 38} 733-840


\bibitem{Fodoretal} Fodor G, Hoenselaers C and Perj{\' e}s Z 1989
{\it J. Math. Phys.} {\bf 30} 2252-2257


\bibitem{Ryan} Ryan F 1995 {\it Phys. Rev. D}  {\bf 52} 5707-5718


\bibitem{derKlis} van der Klis M  2006 {\it Compact Stellar X-Ray
Sources} ed Lewis W H G and van der Klis M  (Cambridge Univ. Press) p 39

\bibitem{stella1} Stella L, Vietri M 1998 {\it ApJ} {\bf 492} L59

\bibitem{pappasQPOs} Pappas G 2012 {\it MNRAS} {\bf 422} 2581

\bibitem{LaarPois} Laarakkers W G, Poisson E 1999 {\it ApJ} {\bf 512} 282-287


\bibitem{Sotiriou}Sotiriou T P, Pappas G 2005  {\it J. Phys.: Conf. Series} {\bf 8} 23

\bibitem{Pachon} Pach{\'o}n L A, Rueda J A, Sanabria-G{\'o}mez J D 2006 {\it Phys. Rev. D} {\bf 73} 104038

\bibitem{Pappas2} Pappas G 2009 {\it J. Phys.: Conf. Series} {\bf 189} 012028

\bibitem{PachonNew} Pach\'{o}n L A, Rueda J A, Valenzuela-Toledo C A 2012 {\it ApJ} {\bf 756} 82

\bibitem{pappas2sol} Pappas G, Apostolatos T A {\it Preprint} arXiv:1209.6148 [gr-qc]

\bibitem{HT} Hartle J B, Thorne K S 1968 {\it ApJ} {\bf 153} 807

\bibitem{Ruffini} Boshkayev K, Quevedo H, Ruffini R 2012 {\it Physical Review D} {\bf 86} 064043

\bibitem{PappasMoments} Pappas G, Apostolatos T A 2012 {\it Phys. Rev. Lett.} {\bf 108} 231104

\bibitem{manko1} Manko V S, Mielke E W, Sanabria-G\'omez J D 2000 {\it Phys. Rev.} D {\bf 61} 081501

\bibitem{manko2} Manko V S, Sanabria-G\'omez J D and Manko O V  2000 {\it Phys. Rev.} D {\bf 62}
044048


\bibitem{SterFrie} Stergioulas N and Friedman J L 1995 {\it ApJ } {\bf 444} 306;\\
Stergioulas N, "rns", (November, 1997), [public domain code]: cited on 19
November 1997, http://www.gravity.phys.uwm.edu/rns


\bibitem{ButtIpse} Butterworth E M and Ipser J R  1976 {\it ApJ} {\bf 204} 200-223


\bibitem{KEH} Komatsu H, Eriguchi Y and Hechisu I 1989 {\it MNRAS} {\bf 237} 355-379


\bibitem{CST} Cook G B, Shapiro S L and Teukolsky S A 1994 {\it ApJ} {\bf 424} 823-845


\bibitem{BertSter} Berti E and Stergioulas N 2004 {\it MNRAS} {\bf 350} 1416


\bibitem{twosoliton} Manko V S, Martin J, Ruiz J E 1995 {\it J. Math. Phys} {\bf 36} 3063

\end{thebibliography}
\end{document}